\documentstyle[aps,prl]{revtex}

\def\be{\begin{equation}}
\def\bea{\begin{eqnarray}}
\def\bma{\begin{mathletters}}
\def\ee{\end{equation}}
\def\eea{\end{eqnarray}}
\def\ema{\end{mathletters}}

\begin{document}
\author{Vlatko Vedral}
\title{High Temperature Macroscopic Entanglement}
\address{Optics Section, Blackett Laboratory, Imperial College London,\\
Prince Consort Road SW7 2BZ, London, United Kingdom}
\date{\today}
\maketitle

\begin{abstract}
In this paper I intend to show that macroscopic entanglement is possible at high
temperatures. I analyze multipartite entanglement produced by the $\eta$ pairing
mechanism which features strongly in the fermionic lattice models of high $T_c$
superconductivity. This problem is shown to be equivalent to calculating multipartite
entanglement in totally symmetric states of qubits. I demonstrate that we can
conclusively calculate the relative entropy of entanglement within any subset of
qubits in an overall symmetric state. Three main results then follow. First, I show
that the condition for superconductivity, namely the existence of the off diagonal
long range order (ODLRO), is not dependent on two-site entanglement, but on just
classical correlations as the sites become more and more distant. Secondly, the
entanglement that does survive in the thermodynamical limit is the entanglement of the
total lattice and, at half filling, it scales with the log of the number of sites. It
is this entanglement that will exist at temperatures below the superconducting
critical temperature, which can currently be as high as $160$ Kelvin. Thirdly, I prove
that a complete mixture of symmetric states does not contain any entanglement in the
macroscopic limit. On the other hand, the same mixture of symmetric states possesses
the same two qubit entanglement features as the pure states involved, in the sense
that the mixing does not destroy entanglement for finite number of qubits, albeit it
does decrease it. Maximal mixing of symmetric states also does not destroy ODLRO and
classical correlations. I discuss various other inequalities between different
entanglements as well as generalizations to the subsystems of any dimensionality (i.e.
higher than spin half).

\end{abstract}

\section{Introduction}

Entanglement is currently one of the most researched phenomena in physics.  Often
shrouded in mystery, its basic premise is quite simple - entanglement is a correlation
between distant particles that exists outside of any description offered by classical
physics. Whilst this may at first glance seem an innocuous statement, in reality it is
anything but. Predictions from the theory of entanglement have confounded some of the
greatest minds in science. Einstein famously dubbed it spukhafte Fernwirkungen:
``spooky action at a distance". As we look deeper into the fabric of nature this
``spooky" connection between particles is appearing everywhere, and its consequences
are affecting the very (macroscopic) world that we experience. At an implementational
level, using entanglement researchers have succeeded in teleporting information
between two parties, designing cryptographic systems that cannot be broken and
speeding up computations that would classically take a much longer time to execute
\cite{Nielsen}. Even though these applications have generated significant interest, I
believe we have only scratched the ``tip of the iceberg" in terms of what entanglement
is, and indeed what we can do with it.

Whilst entanglement is experimentally pretty much beyond dispute in microscopic
systems - such as two photons or two atoms - many people find it difficult to accept
that this phenomenon can exist and even have effects macroscopically. Based on our
everyday intuition we would, for example, find it very hard to believe that two cats
or two human beings can be quantum entangled. Yet quantum physics does not tell us
that there is any limitation to the existence of entanglement. It can, in principle
and as far as we understand, be present in systems of any size and under many
different external conditions.

The usual argument against seeing macroscopic entanglement is that large systems have
a large number of degrees of freedom interacting with the rest of the universe and it
is this interaction that is responsible for destroying entanglement. If we can exactly
tell the state that a system is in, then this system cannot be entangled to any other
system. In everyday life, objects exist at room (or comparable) temperatures so their
overall state is quantum mechanically described by a very mixed state (this mixing due
to temperature is, of course, also due to the interaction with a large ``hot"
environment). Mixing states that are entangled, in general, reduces entanglement and
ultimately all entanglement vanishes if the temperature is high enough. The question
then is how high is the highest temperature before we no longer see any entanglement?
And how large can the body be so that entanglement is still present? Can we, for
example, have macroscopic entanglement at the room temperature?

Entanglement has recently been shown to affect macroscopic properties of solids, such
as its magnetic susceptibility and heat capacity, but at a very low (critical)
temperature \cite{Nature}. This extraordinary result demonstrates that entanglement
can have a significant effect in the macroscopic world. The basic reason for this
dependence is simple. Magnetic susceptibility is proportional to the correlation
between nuclear spins in the solid. As we said before, entanglement offers a higher
degree of correlation than anything allowed by classical physics and the corresponding
quantum susceptibility - which fully agrees with experimental results \cite{Nature} -
is higher than that predicted by using just classical correlations (for further
theoretical support for this see my article in \cite{NJPhys}). It is now very
important to go beyond this low temperature regime and experimentally test
entanglement at higher and higher temperatures.

Thinking that high temperature entanglement is linked with (perhaps even responsible
for) some other high temperature quantum phenomena, such as high temperature
superconductivity, is tempting. After all, superconductivity is a manifestation of the
existence of the off diagonal long range order (ODLRO) \cite{Yang1} which is a form of
correlation that still persists in the thermodynamical (macroscopic) limit. However,
it is not immediately obvious that this correlation contains any quantum entanglement.
My main intention in this paper is to show that it does. This correlation contains
multipartite entanglement between all electron pairs in the superconductor. To
calculate this we need to be able to quantify entanglement exactly and be able to
discriminate entanglement from any form of classical correlation.

A great deal of effort has gone into theoretically understanding and quantifying
entanglement \cite{Vedral1}. There are a large number of different proposed measures;
the different measures capture different aspects of entanglement. In this paper we
will be interested in a measure that is based on the (asymptotic) distinguishability
of entangled states from separable (disentangled) states known as the relative entropy
of entanglement \cite{PRL,Vedral4}. The main advantage of this measure is that it is
easily defined for any number of systems of any dimensionality, which is not the case
for entanglement of formation or distillation \cite{Vedral1}. I have argued that a
number of results in quantum information and computation follow from the relative
entropy function \cite{Vedral1}.

There is, unfortunately, no closed form for the relative entropy of entanglement, but
this measure can still be computed for a large class of relevant states such as the
pure bipartite states, Werner states and many others \cite{Vedral4}. Most recently,
Wei et al \cite{Wei} have succeeded in obtaining a formula for the relative entropy of
entanglement for any number of totally symmetric pure states of $n$ qubits using a
very simple and elegant argument (some partial results have been obtained previously
in this direction using different methods by Plenio and Vedral \cite{Vedral2}, but
only for three qubit symmetric states). I will use and extend these results further
with the idea of applying them to a specific model of a superconductor.

The purpose of this paper is to investigate possible links between high temperature
entanglement and high temperature superconductivity with the intention of showing that
entanglement can persist at higher temperatures. I analyze a particular mechanism -
the $\eta$-pairing of electrons due to Yang \cite{Yang2} - that was originally
proposed to explain high temperature superconductivity. The chief difference between
this pairing mechanism and the usual Bardeen, Cooper and Schrieffer (BCS) electron
pairing \cite{BCS} for (low temperature) superconductivity is that, in the former,
electrons that are positioned at the same site are paired, while in the latter,
electrons forming Cooper pairs are separated by a certain finite average distance (the
so called coherence length, typically of the order of hundreds of nanometers). The
physical reason behind electron pairing is also thought to be different in a high
temperature superconductor, but I do not wish to enter into discussing these details
here (see e.g. \cite{Plakida}). I will, however, look at the $\eta$ model in a
different way, using totally symmetric states, and this will make calculating
entanglement easier. Wei et al. \cite{Wei} have recently made very important steps in
calculating the relative entropy of entanglement for symmetric state. I extend their
approach to calculating the relative entropy of entanglement for mixed symmetric state
arising from tracing over some qubits in pure states, and apply it to understanding
various relations between entanglements of a subset of qubits and their relation to
the total entanglement. I show that although two-site entanglement disappears as the
distance between sites diverges (a conclusion also reached by Zanardi and Wang in a
different way \cite{Zanardi}), the total entanglement still persists in the
thermodynamical limit. Furthermore, it scales logarithmically with the number of
qubits. Therefore, it is this total entanglement that should be compared with ODLRO
and not the two-site entanglement. While the two-site entanglement vanishes
thermodynamically, two-site classical correlations are still present and so is the
entanglement between two clusters of qubits (two cluster entanglement in $\eta$ states
has also been analysed by Fan \cite{Fan}). I show that all aspects of my analysis can
easily be generalised to higher than half spin systems. My hope is that this work -
which is really just a first step in exploring high temperature entanglement - will be
extended to different models with states other than symmetric and that this will allow
us a much more complete understanding of entanglement and the role it plays in the
macroscopic world.

\section{$\eta$-pairing in Superconductivity}

The model that I describe now consists of a number of lattice sites, each of which can
be occupied by fermions having spin up or spin down internal states.  Let us introduce
fermion creation and annihilation operators, $c^{\dagger}_{i,s}$ and $c_{i,s}$
respectively, where the subscript $i$ refers to the $i$th lattice site and $s$ refers
for the value of the spin, $\uparrow$ or $\downarrow$. Since fermions obey the Pauli
exclusion principle, we can have at most two fermions attached to one and the same
site. The $c$ operators therefore satisfy the anticommutation relations:
\begin{equation}
\{c_{i,s},c^{\dagger}_{j,t}\} = \delta_{ij}\delta_{s,t}
\end{equation}
and $c$'s and $c^{\dagger}$'s anticommute as usual. (Some general features of
fermionic entanglement - arising mainly from the Pauli exclusion principle - have been
analysed in \cite{Zanardi,Vedral3,rest,Schliemann}).

We only need assume that our model has the interaction which favors formation of
Cooper pairs of fermions of opposite spin at each site \cite{Yang2}. The actual
Hamiltonian is not relevant for my present purposes. It suffices to say that Yang
originally considered the Hubbrad model for which the $\eta$ states are eigenstates
(but none of them is a ground state \cite{Yang2}). A generalisation of the Hubbard
model was presented in \cite{Korepin} and in a specific regime of this new model the
$\eta$ states do become lowest energy eigenstates (this is a fact that will become
relevant when we talk about high temperature entanglement). Both these models have
been used to simulate high-temperature superconductivity, since in high
superconducting materials, the coherent length of each Cooper pair is on average much
smaller than for a normal superconductor.

Suppose, now, that there are $n$ sites and suppose, further, that we introduce an
operator $\eta^{\dagger}$ that creates a coherent superposition of a Cooper pair in
each of the lattice sites,
\begin{equation}
\eta^{\dagger} = \sum_{i=1}^n c^{\dagger}_{i,\uparrow} c^{\dagger}_{i,\downarrow} \; .
\end{equation}
The $\eta^{\dagger}$ operator can be applied to the vacuum a number of times, each
time creating a new coherent superposition. However, the number of applications, $k$,
cannot exceed the number of sites, $n$, since we cannot have more than one pair per
site due to the exclusion principle. I now introduce the following basis
\begin{equation}
|k,n-k\rangle := \frac{1}{\sqrt{n \choose k}} (\eta^{\dagger})^k |0\rangle \; ,
\end{equation}
where the factor in front is just the necessary normalisation. Here, the vacuum state
$|0\rangle$ is annihilated by all $c$ operators, $c_{i,s} |0\rangle = 0$. We note in
passing that the originally defined $\eta$ operators can also have phase factors
dependent on the location of the site on the lattice. We can have a set of operators
like
\begin{equation}
\eta_k = \sum_n e^{ikn} c^{\dagger}_{n,\uparrow} c^{\dagger}_{n,\downarrow} \; .
\end{equation}
All the states generated with any $\eta_k$ from the vacuum will be shown to have the
same amount of entanglement so that the extra phases will be ignored in the rest of
the paper (i.e. we will only consider the $k=0$ states).

We can think of the $\eta$ states in the following way. Suppose that $k=2$. Then this
means that we will be creating two $\eta$-pairs in total, but they cannot be created
in the same lattice site. The state $|2,n-2\rangle$ is therefore a symmetric
superposition of all combinations of creating two pairs at two different sites. Let
us, for the moment, use the label $0$ when the site is unoccupied and $1$ when it is
occupied. Then the state $|2,n-2\rangle$ is
\begin{equation}
|2,n-2\rangle =\frac{1}{\sqrt{n \choose 2}}
(|\underbrace{000}_{n-2}...\underbrace{11}_{2}\rangle +
...|\underbrace{11}_{2}...\underbrace{000}_{n-2}\rangle)
\end{equation}
i.e. it is an equal superposition of states containing $2$ states $|1\rangle$ and
$n-2$ states $|0\rangle$. These states, due to their high degree of symmetry, are much
easier to handle than general arbitrary superpositions and we can compute entanglement
for them between any number of sites. Note that in this description each site
effectively holds one quantum bit, whose $0$ signifies that the site is empty and $1$
signifies that the site is full.

The main characteristic of $\eta$ states is the existence of the long range off
diagonal order (ODLRO), which implies its various superconducting features, such as
the Meissner effect and flux quantisation \cite{Nieh}. The ODLRO is defined by the off
diagonal matrix elements of the two-site reduced density matrix being finite in the
limit when the distance between the sites diverges. Namely,
\begin{equation}
\lim_{|i-j|\rightarrow \infty} \langle c^{\dagger}_{j,\uparrow}
c^{\dagger}_{j,\downarrow} c_{i,\downarrow} c_{i,\uparrow} \rangle \longrightarrow
\alpha \label{ODLRO}
\end{equation}
where $\alpha$ is a constant (independent of $n$). I will show that although the
existence of off diagonal matrix elements does not guarantee the existence of
entanglement between the two sites, it does guarantee the existence of multi-site
entanglement between all the sites. Note that here, by ``correlations" I mean
correlations between the number of electrons positioned at different sites $i$ and
$j$. Namely, we are looking at the probability of one site being occupied (empty)
given that the other site is occupied (empty). This is different from spin-spin
correlations, which would look at the occurrences of both electron spins being up or
down, or one being up and the other being down \cite{Vedral3}.

\section{General Description of Symmetric States}

The states I will analyze here will always be of the form
\begin{equation}
|\Psi (n,k)\rangle \equiv |k, n-k\rangle := \frac{1}{\sqrt{n \choose
k}}(\hat{S}|\underbrace{000}_{k}...\underbrace{11}_{n-k}\rangle)
\end{equation}
where $\hat{S}$ is the total symmetrisation operator. We will also consider mixtures
of these states, which become relevant when we talk about systems at finite
temperatures. Symmetric states arise, for example, in the Dicke model in which $n$
atoms simultaneously interact with a single mode of the electro-magnetic field
\cite{Dicke}. They are, furthermore, very important as they happen to
be eigenstates of many models in solid state physics, and, in particular, they are
eigenstates of the Hubbard and related models supporting the $\eta$ pairing mechanism.
The analysis presented in this paper will be applicable to any of these systems and
not just the $\eta$ model. The $\eta$ mechanism will be here significant because of
its potential to support high temperature entanglement.

I would now like to start to compute the entanglement between every pair of qubits
(sites) in the above state $|\Psi (n,k)\rangle$. A simpler task would be first to tell
if and when every pair of qubits in a totally symmetric state is entangled. For this,
we need only compute the reduced two-qubit density matrix which can be written as:
\begin{equation}
\sigma_{12}(k) = a |00\rangle \langle 00| + b |11\rangle \langle 11| + 2c
|\psi^{+}\rangle \langle \psi^{+}|
\end{equation}
where $|\psi^{+}\rangle = (|01\rangle + |10\rangle)/\sqrt{2}$ and
\begin{eqnarray}
a & = & \frac{{n-2 \choose k-2}}{{n \choose k}} = \frac{k(k-1)}{n(n-1)}\\
b & = & \frac{{n-2 \choose k}}{{n \choose k}} = \frac{(n-k)(n-k-1)}{n(n-1)}\\
c & = & \frac{{n-2 \choose k-1}}{{n \choose k}} = \frac{k(n-k)}{n(n-1)} \; .
\end{eqnarray}
We can easily check that $a+b+2c =1$ and so the state is normalized. This density
matrix is the same no matter how far the two sites are from each other, since the
state is symmetric, and must therefore be identical for all qubits. We can easily test
the Peres-Horodecki (partial transposition) condition \cite{Peres} for separability of
this state. Two states are entangled if and only if they are inseparable which leads
to states $\sigma_{12} (k)$ being entangled if and only if
\begin{equation}
a + b - \sqrt{(a-b)^2 + 4c^2} < 0 \; ,
\end{equation}
which leads to
\begin{equation}
(k-1) (n-k-1) < k (n-k) \; .
\end{equation}
This equation is satisfied for all $n\ge 2$ (two qubits or more) and $1 \le k \le
n-1$. So, apart from the case when the total state is of the form $|000..0\rangle$ or
$|111..1\rangle$, there is always two-qubit entanglement present in symmetric states.
Note, however, that in the limit of $n$ and $k$ becoming large - no matter what their
ratio may be - the value of the left hand side approaches the value of the right hand
side and entanglement thus disappears. This is a very interesting property of
symmetric states and we will be able to quantify it exactly in the next section.

An important point to make is that the two point correlation function used in the
calculation of the ODLRO in eq. (\ref{ODLRO}) is, in fact, just one of the sixteen
numbers we need for the full two-site density matrix (the independent number of real
parameters is actually fifteen, because of normalisation). In our simplified case of
symmetric states in the $\eta$-pairing model, this off diagonal element is equal to
$c$. However, for the density matrix we still need to know $a$ and $b$, and these
numbers clearly affect the amount of entanglement. Imagine, for example, the situation
where $a=b$. Then the condition for entanglement is that $a-c <0$, which does not hold
if $a\ge c$ and such a density matrix is certainly possible. So, the first lesson is
that two-site entanglement is not the same as the existence of ODLRO, and therefore
two-site entanglement is not relevant for superconductivity. This does not mean, of
course, that there is no entanglement in the whole of the lattice. In the next
section, I will calculate exactly this. We will determine the relative entropy of
entanglement for all symmetric states and all their substates. I will be able to
extend the method of Wei et al \cite{Wei} and analyze many relationships between
various subsets of symmetric states, including the amount of entanglement in any
subset of qubits (or sites).

\section{Relative Entropy of Entanglement for Symmetric States}

The symmetric states are very convenient for studying various features of multipartite
entanglement simply because, as we already indicated, we can compute exactly the
relative entropy of entanglement for any reduced state including the total symmetric
state for any $n$ and $k$. It is expected that, because they possess a high degree of
symmetry, they will also display a high degree of entanglement. It is precisely for
this reason that they are suitable to allow the existence of entanglement at high
temperatures. This will now be analyzed in detail.

I first introduce the relative entropy of entanglement. The relative entropy of
entanglement measures the distance between a state and the nearest disentangled
(separable) state. If ${\cal D}$ is the set of all disentangled states (i.e. states of
the form $\sum_i p_i \rho^i_1 \otimes \rho^i_2\ldots\otimes \rho^i_n$, where $p_i$ is
any probability distribution), the measure of entanglement for a state $\sigma$ is
then defined as
\begin{equation}
E({\sigma}):= \min_{\rho \in \cal D}\,\,\, S(\sigma || \rho)\; , \label{measure}
\end{equation}
where $S(\sigma || \rho) = tr (\sigma \log \sigma - \sigma \log \rho)$ is the relative
entropy between the two density matrices $\rho$ and $\sigma$. In order to compute this
measure for any state $\sigma$ we need to be able to find its closest disentangled
state $\rho$. Finding this closest state is, in general, still an open problem,
however, it has recently been solved for pure symmetric states by Wei et al
\cite{Wei}.

Wei et al showed that a convenient and intuitive way of writing the closest
disentangled state to the symmetric state $|k, n-k\rangle$ is \cite{Wei}:
\begin{equation}
\rho = \frac{1}{2\pi} \int_0^{2\pi} d\phi |\phi^{\otimes n}\rangle \langle
\phi^{\otimes n}| \label{close} \; ,
\end{equation}
where
\begin{equation}
|\phi^{\otimes n}\rangle = (\sqrt{k/n} |0\rangle + \sqrt{(n-k)/n}
e^{i\phi} |1\rangle)^{\otimes n}
\end{equation}
is the tensor product of $n$ states each of which is a superposition of the states
$|0\rangle$ and $|1\rangle$ with probabilities $k/n$ and $1-k/n$ respectively. This
$\rho$ was proved to achieve the minimum of the relative entropy by showing that it
saturates an independently obtained lower bound. The relative entropy of entanglement
of the total state is now easily computed. Since $\sigma = |k,n-k\rangle\langle
k,n-k|$ is a pure state, $tr \sigma \log \sigma = 0$ and we only need to compute
$-\langle k,n-k| \log \rho |k,n-k\rangle$, which is equal to
\begin{equation}
E(|k, n-k\rangle) = - \log {n \choose k} + k \log \frac{n}{k} + (n-k) \log
\frac{n}{n-k}\; .
\end{equation}
Note that entanglement is largest when $n=2k$ as is intuitively expected (i.e. the
largest number of terms is then present in the expansion of the state in terms of the
computational basis states). Then, for large $n$, it can be seen that the amount of
entanglement grows as
\begin{equation}
E(|n/2, n/2\rangle) \approx  \frac{1}{2} (\log n + 2)
\end{equation}
and so (in the leading order) entanglement grows logarithmically with the number of
qubits in the state. To obtain this formula I have used Sterling's approximation for
the factorial
\begin{equation}
n! \approx 2.507 n^{n+1/2}e^{-n} \; .
\end{equation}
Most results in this paper will asymptotically have the form $\alpha \log n + \beta$ where
$\alpha >0$ and $\beta$ are constants that will usually be omitted as we only care about
the general form of the behaviour.

I now return to the question of different phases introduced between different elements of the
superposition in the symmetric states. Let us consider states of the form
\begin{equation}
|1,n-1,\theta\rangle = |00..1\rangle + e^{i\theta} |00..10\rangle + e^{(n-1)i\theta}
|10..0\rangle \; ,
\end{equation}
where we have $k=1$ ones and $n-1$ zeroes and $\theta$ is any phase. The simplest way
of seeing that entanglement does not depend on the phase $\theta$ is to define a new
basis at the $m$th site as $|\tilde{0}\rangle = |0\rangle , |\tilde{1}\rangle = \exp
\{(m-1) \theta\} |1\rangle $. This way the phases have been absorbed by into the basis
states and the resulting state is, in the tilde basis,
\begin{equation}
|1,n-1,\theta\rangle = |\tilde{0}\tilde{0}..\tilde{1}\rangle +
|\tilde{0}\tilde{0}..\tilde{1}\tilde{0}\rangle +
|\tilde{1}\tilde{0}..\tilde{0}\rangle\; .
\end{equation}
The amount of entanglement must therefore be independent of any phase difference of
the above type and this is, of course, true for symmetric states with any number of
zeroes and ones. All considerations from this point onwards will therefore immediately
apply to all these states will different phases.

We can also compute the two-site relative entropy of entanglement exactly. The closest
disentangled state is in this case the same as in eq. (\ref{close}) with $n=2$. In the
computational basis we have
\begin{equation}
\rho = \left(\frac{k}{n}\right)^2 |00\rangle \langle 00| +
\left(\frac{n-k}{n}\right)^2 |11\rangle \langle 11| + \left(\frac{2k(n-k)}{n^2}\right)
|\psi^{+}\rangle \langle \psi^{+}|\; . \label{closest}
\end{equation}
That this is a minimum can be seen from the fact that the relative entropy of the
state of two qubits is:
\begin{eqnarray}
S(\sigma ||\rho) & = & -S(\sigma) - \langle \psi^{+} | \log \rho
|\psi^{+}\rangle - \langle 00 | \log \rho |00\rangle  - \langle 11
| \log \rho |11\rangle \\
&\ge & -S(\sigma) - \log \langle \psi^{+} | \rho |\psi^{+}\rangle - \log \langle 00 |
\rho |00\rangle  - \log \langle 11 | \rho |11\rangle \; ,
\end{eqnarray}
the inequality following from concavity of the $\log$ function. Suppose now that
$\rho$'s only non-zero elements are $\rho_{00}=\langle 00 | \rho
|00\rangle,\rho_{11}=\langle 11 | \rho |11\rangle$ and $\rho_{++} = \langle \psi^{+} |
\rho |\psi^{+}\rangle$. Given that it has to be separable, meaning that
$2\sqrt{\rho_{00}\rho_{11}}\ge \rho_{++}$ (which follows from the Peres-Horodecki
criterion), and that, at the same time, it has to be closest to $\sigma$, we can
conclude that $\rho_{00} =k/n$. The other entries of $\rho$ then follow.

To prove that $\rho$ is the minimum in a rigorous fashion, we need to show that any
variation of the type $(1-x)\rho + x \omega$ where $\omega$ is any separable state
leads to a higher relative entropy (a method similar to \cite{Vedral4}). Since
relative entropy is a convex function, this means that
\begin{equation}
\frac{d}{dx} S(\sigma ||(1-x)\rho + x \omega) \ge 0 \; .
\end{equation}
In fact, since relative entropy is convex in the second argument it is enough to
assume that $\omega$ is just a product state.

For $a > 0$, $ \log a = \int_0^\infty {at - 1\over a + t} {dt \over 1 + t^2}$, and
thus, for any positive operator $A$, $ \log A = \int_0^\infty {At - 1\over A + t} {dt
\over 1 + t^2}$. Let $f(x, \omega) = S(\sigma||(1-x) \rho + x \omega)$. Then
\begin{eqnarray}
 {\partial f \over \partial x}(0, \omega) & = & -\lim_{x \rightarrow 0}
\mbox{Tr}\bigg \{{\sigma (\log ((1-x) \rho + x \omega) -
\log \rho)\over x}\bigg \} \nonumber \\
&  =  & \mbox{Tr}\bigg \{ ( \sigma \int_0^\infty (\rho +
t)^{-1} ( \rho - \omega) (\rho +t)^{-1} dt \big) \bigg \} \nonumber  \\
& = & 1 -\int_0^\infty \mbox{Tr}( \sigma (\rho + t)^{-1}
\omega  (\rho +t)^{-1} \big) dt \nonumber \\
& = & 1 - \int_0^\infty \mbox{Tr}( (\rho + t)^{-1} \sigma (\rho + t)^{-1} \omega \big)
dt \; .
\end{eqnarray}
For our minimal guess $\rho$ in eq. (\ref{closest}) we can then write
\begin{eqnarray}
{\partial f \over \partial x}(0, \omega) - 1 & = & - \mbox{Tr} \bigg \{ \omega
\int_0^\infty (\rho + t)^{-1} \sigma (\rho + t)^{-1} dt  \bigg \} \nonumber
\\ & = & \frac{n}{n-1} \frac{k-1}{k} \langle 00|\omega|00\rangle +
\frac{n}{n-1} \frac{n-k-1}{n-k} \langle 11|\omega|11\rangle \nonumber \\
& + & \frac{n}{n-1} \langle \psi^+|\omega|\psi^+\rangle \; ,
\end{eqnarray}
where we have used the fact that $\int_0^\infty (p + t)^{-2} dt = p^{-1}$. Since the
expression in the previous equation is always less than or equal to a unity if $\omega
= |\alpha\beta\rangle\langle\alpha\beta|$ (i.e. a product state), it follows that
\begin{eqnarray}
\left|{\partial f \over \partial x}(0, \omega) - 1\right| & \leq & 1 \;\; .
\end{eqnarray}
Thus it also follows that ${\partial f \over \partial x} (0,
|\alpha\beta\rangle\langle\alpha\beta|) \geq 0$. But any separable state can be
written in the form $\rho = \sum_i r_i |\alpha^i\beta^i\rangle\langle\alpha^i\beta^i|$
and so
\begin{equation}
{\partial f \over \partial x}(0, \rho) = \sum_i r_i {\partial f \over
\partial x}(0, |\alpha^i\beta^i\rangle\langle\alpha^i\beta^i|) \geq 0 \; .
\end{equation}
And this confirms that $\rho$ is the minimum since the gradient is positive.

Therefore the relative entropy of entanglement between any two sites is:
\begin{eqnarray}
E_{12} & = & a\log a - b\log b - 2c\log 2c \nonumber \\
& - & a \log \left(\frac{k}{n}\right)^2 - b \log \left(\frac{n - k}{n}\right)^2 - 2c
\log \left(\frac{2k(n-k)}{n^2}\right) \nonumber \\
& = & \log\left( \frac{n}{n-1} \right)+ \frac{k(k-1)}{n(n-1)} \log \left(
\frac{k-1}{k}\right) + \frac{(n-k)(n-k-1)}{n(n-1)} \log
\left(\frac{n-k-1}{n-k}\right)\; .
\end{eqnarray}
We see that when $n,k,n-k \rightarrow \infty$, then $E_{12} \rightarrow 0$ as it
should be from our discussion of the separability criterion. This can be thought of as
one way of recovering the ``quantum to classical" correspondence in the limit of large
number of systems present in the state: locally, between any two sites, entanglement
does vanish, although globally, and as will be seen in more detail, entanglement still
persists.

Entanglement of any number of qubits, $l\le k$, can also be calculated using the same
method . The state after we trace out all but $l$ qubits is given by
\begin{equation}
\sigma_l = \sum_{i=0}^l {l\choose l-i} \frac{{n-l \choose k-i}}{{n \choose k}}
|i,l-i\rangle\langle i,l-i| \; .
\end{equation}
The closest disentangled state is given by
\begin{equation}
\rho_l = \sum_{i=0}^l {l\choose i} \left(\frac{k}{n}\right)^{l-i}
\left(\frac{n-k}{n}\right)^{i} |i,l-i\rangle\langle i,l-i| \; ,
\end{equation}
as can be shown by the above method. The relative entropy of entanglement is now given
by
\begin{equation}
E_l = \sum_{i=0}^l {l\choose l-i} \frac{{n-l \choose k-i}}{{n \choose k}} \log \left\{
{l\choose l-i}\frac{{n-l \choose k-i}}{{n \choose k}}
(\frac{n}{k})^{l-i}(\frac{n}{n-k})^{i} {l \choose i}^{-1}\right\} \label{lqubitent}\;
.
\end{equation}
This is a very interesting quantity as it allows us to speak about entanglement
involving any number of qubits. What do we expect from it? We expect that entanglement
grows exponentially with $l$, for a fixed total number of qubits, $n$. This can be
confirmed using the Sterling formula. Note that entanglement grows at this rate even
though the states we are talking about are mixed, since $n-l$ qubits have been traced
out. Another way of seeing why entanglement grows exponentially with the number of
qubits included for a total fixed number of qubits, is to look at the opposite regime.
For any finite fixed $l$, we should have that in the large $n,k,n-k$ limit the amount
of entanglement between $l$ tends to zero. This decrease with larger and larger $n$
happens at an exponential rate.

\section{Classical Versus Quantum Correlations}

In this section I would like to investigate the relationship between classical and
quantum correlations for symmetric states, and both in relation to the already
introduced concept of ODLRO. First of all, it is clear that in the limit of
$n\rightarrow \infty$ all bipartite (or two-site) entanglement disappears (this was
seen both from the Peres-Horodecki criterion and from the direct computation of the
relative entropy). In spite of this, the ODLRO still exists and the two quantities are
therefore not related. In other words, two-site entanglement is not relevant for
superconductivity. However the main point of this section is that the two-site
classical correlations still survive in the limit of $n\rightarrow \infty$. In order
to show this, let us, first of all, define bipartite classical correlations.

A quantum state can have zero amount of entanglement, but still have non-zero
classical correlations. An example is the state $|00\rangle\langle 00|+
|11\rangle\langle 11|$. Classical correlations between systems $A$ and $B$ in the
state $\sigma_{AB}$ can be defined as \cite{Henderson}
\begin{equation}
C_A (\sigma_{AB}):= \max_{A_i^{\dagger}A_i} S(\sigma_B) - \sum_i p_i
S(\sigma_B^i)=\max_{A_i^{\dagger}A_i}\sum_i p_i S(\sigma_B^i||\sigma_B)\; ,
\end{equation}
where $\sigma_B^i=tr_A \sigma_{AB}^i$, $\sigma_{AB}^i=A_i \sigma_{AB}A_i^{\dagger}$,
and $\sum_i A_i^{\dagger}A_i=1$ is the most general measurement on system $A$. The
same can be defined with the most general measurement performed on $B$, so that we
obtain
\begin{equation}
C_B (\sigma_{AB}): = \max_{B_i^{\dagger}B_i} S(\sigma_A) - \sum_i p_i
S(\sigma_A^i)=\max_{B_i^{\dagger}B_i}\sum_i p_i S(\sigma_A^i||\sigma_A) \; .
\end{equation}
The physical motivation behind the above definition is the following: classical
correlations between $A$ and $B$ tell us how much information we can obtain about $A$
($B$) by performing measurements in $B$ ($A$). It is the (maximum) difference between
the entropy of $A$ ($B$) before and after the measurement on $B$ ($A$) is performed.
There is some evidence that $C_A=C_B$ \cite{Henderson}, but this equality will not be
relevant here.

Now, applying this measure of classical correlations to the two-site reduced density
matrix from the overall symmetric state, $\rho_{12}$, we obtain,
\begin{eqnarray}
C & = & -a\log a - b\log b - c\log c + \frac{1}{2} ((a+c/2)\log (a+c/2) + (b+c/2)\log
(b+c/2)) \nonumber\\
& = & (r-2r^2) \log r + ((1-r) - 2(1-r)^2) \log (1-r) - 2r(1-r) \log 2r(1-r)
\end{eqnarray}
where $r=k/n$ is the fraction of ones in the state (the so called filling factor in
any ``Cooper pair" lattice model, including the $\eta$ model). We now see that at half
filling - when ODLRO is maximal - the classical two-site correlations also survive
asymptotically since $C_A=C_B=0.5$. Therefore, all the correlations between any two
sites are here due to classical correlations.

Note, incidentally,  that we cannot have the situation in which entanglement exists
between two parties, while at the same time classical correlations vanish. Quantum
correlations presuppose the existence of classical correlations. This, of course,
relies on the fact that entanglement is defined in a reasonable way, namely that when
we talk about two-site entanglement we must trace the other sites out. We are not
allowed to perform measurements on other sites and condition the remaining
entanglement on them. Measurements that generate entanglement are, first of all,
unrealistic for a macroscopic object which thermalizes very quickly. Even if we were
to allow such measurements, then the state after them will still have classical
correlations of at least the same magnitude as entanglement. So, it cannot be that
entanglement is important for the issues of superconductivity, phase transitions,
condensation, etc., and that classical correlations are not.

As an example, let us take the ``maxmum singlet fraction" in the two-site density
matrix $\sigma_{12}$ as our definition of entanglement. This is the maximaum fraction
of a maximally entangled state in the state $\sigma_{12}$, which is in this case equal
to $c$, and this is the same as ODLRO. So, if the maximum singlet fraction is used to
measure entanglement, then entanglement also persists in the thermodynamical limit. In
fact, as will be shown later, this measure also survives when we mix symmetric states,
because it is a linear measure. The maximum singlet fraction, however, is not a
realistic measure of entanglement as it is not easily accessible experimentally, which
is why we do not use it in this paper.

In order to make our analysis more complete we also show how to calculate mutual
information \cite{Vedral1} for symmetric states. This quantity tells us about the
total (quantum plus classical) correlations in a give state. Mutual information is
equal to the relative entropy between the state itself and the product of individual
qubit density matrices, obtained by tracing out all the other qubits. This product
state is easily written down to be:
\begin{equation}
\rho_{prod} = \left(\frac{k}{n} |0\rangle\langle 0| + \frac{n-k}{n} |1\rangle\langle
1|\right)^{\otimes n}\; .
\end{equation}
The mutual information is now given by
\begin{equation}
I (|k,n-k\rangle) = n \left(-\frac{k}{n}\log \frac{k}{n} - \frac{n-k}{n} \log
\frac{n-k}{n}\right)\; , \label{mutual}
\end{equation}
and this is basically just the sum of individual qubit entropies. Since the qubit
entropy (the quantity in brackets in the above equation) is a finite quantity for a
given ratio $r=k/n$, the total mutual information grows linearly with the number of
qubits $n$. Furthermore, since entanglement grows as $\log n$, we conclude that
classical correlations grow roughly as $n-\log n$ (for this conclusion to be exact,
classical and quantum correlations as defined here would have to add up to mutual
information; while this is true for some states \cite{Henderson}, it is certainly not
true in general).

The fact that classical correlations and mutual information survive the
thermodynamical limit does not imply that there in no meaning left for entanglement
when it comes to superconductivity and ODLRO. Only now, we must talk either about the
bipartite entanglement between two clusters of sites (to be computed in the next
section) or the multipartite entanglement between all sites. Since the overall state
across all sites is pure in our considerations so far, this means that two-site
non-vanishing classical correlations (or equivalently ODLRO) must imply entanglement
between two clusters, each of which contains one of the sites and such that the union
of the two clusters is the whole lattice. This simply must be the case, since,
otherwise, if the clusters were not entangled, the total state would be a product of
the states of individual clusters, and this means that even classical correlations
would be zero, which is a contradiction. Furthermore, the fact that any two such
clusters are entangled, must mean that the multipartite entanglement also exists,
since this entanglement is by definition larger than any bipartite entanglement (as,
for multipartite entanglement, we are looking for the closest separable state over all
sites, rather than just over the two clusters). I now quantify these various relations
a bit more precisely.

\section{Various Other Relations Between Entanglements}

In this section I will discuss some other results that can be derived from our
knowledge of symmetric states so far. Some of the results will not necessarily be
directly relevant for the main theme of the paper - high temperature entanglement -
but this section is a natural place to present them. The discussion about high
temperature entanglement in the rest of the paper can be understood without reading
this section. I will fully return to the main topic in the next section. The first
important question to be addressed here is the following. Suppose we look at the
entanglement between one qubits and the rest $n-1$ qubits in total and individually.
We would expect that the total one-versus-rest entanglement is larger then the
individual sum of the two-qubit entanglements. The logic behind this conclusion is
that by looking at entanglements individually we always lose something from the total
entanglement, i.e. the operation of tracing reduces entanglement. This translates into
the following inequality:
\begin{equation}
(n-1) E_{12} \le E_{1:(2,3..n)} \; ,
\end{equation}
where
\begin{equation}
E_{1:(2,3..n)} = -\frac{k}{n} \log \left( \frac{k}{n}\right) - \frac{n-k}{n} \log
\left(\frac{n-k}{n}\right)
\end{equation}
is basically the same as the entropy of every qubit in the symmetric state. We can
prove this inequality by noting that it holds for $k=1$ and $2k=n$ (the extreme
points), and because of the monotonicity and continuity of both sides it has to hold
in general.

The aforementioned inequality has a very important implication which shows that the
bipartite entanglement in the symmetric state is always bounded from above by
\begin{equation}
E_{12} \le \frac{E_{1:(2,3..n)}}{(n-1)} \le \frac{1}{n-1} \approx
\frac{1}{n}
\end{equation}
the second inequality following from the fact that the entanglement between one qubit
and the rest is equal to the entropy of that qubit and that can at most be $\log 2 =
1$. Therefore, while the total entanglement of the symmetric state increases with
$\log n$, the two qubit entanglement decreases as $1/n$. Here we see most directly how
it is possible to have the emergence of (only) classical correlations between
constituents even though globally entanglement increases. There are many other open
questions related to this one. We can repeat the same calculation for any fixed number
of qubits. We can check if a cluster of qubits is, for example, more entangled to
another cluster of qubits in total or if we add all the entanglements between their
individual elements. Some of these may not be easy questions to answer in general.

I would now like to calculate the entanglement between $l$ qubits and the remaining
$n-l$ qubits. Since the whole state that we are now examining is pure, the relative
entropy of entanglement is given by the entropy of the $l$ qubits:
\begin{equation}
S_{12...l} = -\sum_{i=0}^l {l\choose l-i} \frac{{n-l \choose k-i}}{{n \choose k}}
\log\left\{ {l\choose l-i} \frac{{n-l \choose k-i}}{{n \choose k}} \right\} \; .
\end{equation}
What are the properties of this expression when we take the various asymptotic limits?
How is this quantity related to other entanglements calculated here? We expect that
for the half filling, $n/k=2$, and $n,l\rightarrow \infty$, the entropy becomes $\log
l$, since we basically have a maximal mixture in the symmetric subspace of $l$ qubits.
This can be confirmed by a simple application of the Sterling approximation formula
used before. The result is in agreement with the fact that total entanglement grows at
the rate of the log of the number of qubits, since two cluster entanglement is a lower
bound for the total entanglement in the state between all the qubits.

The last question I address is the relationship between the lower and higher order
entanglement in the symmetric states. More precisely, the question is: if we add all
the entanglements up to and including $m$ qubits, is this quantity larger or smaller
than the amount of entanglement of $m+1$ qubits? Mathematically, this translates into
the following two possible inequalities:
\begin{equation}
\sum_{i=1}^m E_i \le E_{m+1} \; \mbox{or} \; \sum_{i=1}^m E_i \ge E_{m+1}
\end{equation}
where $E_m$ is given in eq. (\ref{lqubitent}). We already know that for $n=3$ and
$k=1,2$, and $l=3$ we have the equality in the above, namely $E_3 = E_1 + E_2$
\cite{Vedral2}. From this result alone it is not clear which way to expect the
inequality to be. Numerical examples show us that, in fact, both results are possible.
If we check the inequality for $n=100, k=50$ and $l=4$ for example, than the left hand
side is smaller than the right hand side and the first inequality holds. For $n=100, k=50$ and $l=30$,
on the other hand, the left hand side is larger than the right hand side and the second
inequality is satisfied. It is an interesting and open question to investigate the point of
the cross-over when the two sides become equal to each other.

\section{Thermal Entanglement and Superconductivity}

There is a critical temperature beyond which any superconductor becomes a normal
conductor. The basic idea behind computing this temperature according to BCS is the
following. At a very low temperature, only the ground state of the system is populated
and for a superconductor this state involves a collection of Cooper pairs with
different momenta values around the Fermi surface. This state can be, somewhat
loosely, thought of as a Cooper pair condensate, and it is this condensation that is
the key to superconductivity. It took initially a long time to understand how the
pairs are formed, since electrons repel each other and therefore should not be bound
together. The attraction is provided by electrons interacting with the positive ions
left in the lattice. We can think of one electron moving and dragging along the
lattice, which then pulls other electrons thereby providing the necessary attraction
\cite{BCS}. When the temperature starts to increase, the Cooper pairs start to break
up, leading to the transition to the normal conductor. What this ``breaking up" means
is that higher than ground states start to get populated by electrons, and these are
states where an electron is created with say momentum $k$ and spin up, but no electron
is created in the $-k$ momentum state. From the BCS analysis this critical temperature
can be calculated to be of the form \cite{BCS}
\begin{equation}
T_c \approx \frac{\hbar \omega}{k} e^{-1/\lambda}
\end{equation}
where $\hbar \omega$ is the energy shell around the Fermi surface which is engaged in
formation of Cooper pairs, $k$ is the Boltzmann constant and $\lambda$ is a parameter
equal to the product of the electron density at the Fermi surface $N(0)$ and the
effective electronic attractive coupling, $V$. The critical temperature formula is
valid in the weak coupling regime where $\lambda = N(0) V <<1$.

The formula for the critical temperature is usually used for other mechanisms of
electron pairing, and not just coupling via the phonon lattice modes as in the BCS
model \cite{BCS}. Importantly for us, the formula also features in models for
explaining and designing high $T_c$ superconducting materials. If the attraction, say
between an electron and a hole, is of the order of Coulomb forces, $\hbar \omega
\approx 1eV$m, and for the weak coupling of, say, $\lambda = 0.2$, the critical
temperature we obtain is $100K$. So if the material is below this temperature, it is
then superconducting. Anything above $70-90 K$ is considered to be high temperature
superconductivity, since it can be achieved by cooling with liquid nitrogen (which is
a standard and easy method of cooling). What seems to be the mechanism behind high
temperature superconductivity, is the fact that the energy gap between the ground
superconducting, electron-pair state, and the excited states is large enough not to be
easily excited as the temperature increases well beyond zero temperature. The exact
way in which this is achieved is still an open question. In the models mentioned here
the ground state is one of the symmetric states from the previous sections. Therefore,
we can conclude that as long as we have high temperature superconductivity, the total
state should also be macroscopically entangled. Superconduction and hence entanglement
can currently exist at temperatures of about $160$ Kelvin.

I would now like to explicitly calculate and show how entanglement disappears as the
temperature increases for any model having the $\eta$ pairing state as the ground
state. For this, we need to be able to describe other states that would be mixed in
with the symmetric $\eta$ states as the temperature increases. They, of course, depend on
the actual Hamiltonian. For instance, in the Hubbard model in \cite{Yang2}, states of the type
\begin{equation}
\xi^{\dagger}_a |0\rangle = \sum_{i}
c^{\dagger}_{i,\downarrow}c^{\dagger}_{i+a,\uparrow}|0\rangle
\end{equation}
are important; here we create a spin singlet state but at sites separated by the
distance $a\neq 0$. If we have $2k$ electrons in total, then $2k-2$ would be paired in
the lowest energy state, and the remaining two electrons would not be. This would give
us the state of the form
\begin{equation}
|\xi\rangle := \eta^{k-1} \xi^{\dagger}_a |0\rangle \; .
\end{equation}
Note that this state is a symmetric combination of states which have $k-1$ electron
pairs distributed among $n$ sites and the last electron pair is in two different sites
separated by the distance $a$. These two sites are different from the other $k-2$
sites due to Pauli's exclusion principle. Even higher states are obtained by having
two electron pairs existing outside of the symmetric state and so on. The exact form
of these, as noted before, depends on the exact form of the Hamiltonian. Even simple
Hamiltonians are frequently very difficult to diagonalize and their eigenstates are
still by and large unknown. Given this, it may be difficult to calculate the exact
amount of entanglement when, at finite temperature, the ground state is mixed with
higher energy states. I will, therefore, make a simplifying assumption that, if the
ground state is $|k,n-k\rangle$, the higher energy states can be written as
$|k-1,n-k+1\rangle$, $|k-2,n-k+2\rangle$ and so on. All these will in fact be assumed
to be symmetric and I will ignore the extra unpaired electrons as far as entanglement
is concerned (they will only contribute to the eigenvalue of energy as it were).

This assumption leads us to consider mixtures of symmetric states. The symmetric
states will be mixed with probabilities in accordance to Boltzmann's exponential law,
or the Fermi-Dirac law if we talk about $\eta$ pairs. Which distribution we use will
be immaterial for our argument. The total state, $\sigma_T$, is
\begin{equation}
\sigma_T = \sum_{k=0}^n p_k |\Psi (k,n)\rangle \langle \Psi (k,n)|
\end{equation}
where, in the case of $\eta$ pairs, the probabilities are
\begin{equation}
p_i = \frac{1}{e^{E_i/kT} +1}
\end{equation}
where $p_i$ is the probability of occupying the $i$th energy level. The reduced
two-site state can be calculated to be
\begin{equation}
\sigma_{12} = \sum_{k=0}^n p_k \sigma_{12} (k) \; .
\end{equation}
The condition for inseparability now becomes
\begin{equation}
\sum_{k,l} p_{k} p_{l} k(n-l)\{(n-k)l - (k-1)(n-l-1)\} >0 \; .
\end{equation}
We see that the thermal averaging is in a sense inconsequential for the existence of
entanglement as the factors $p_{k}p_{l}$ are probabilities and are always
non-negative. For inequality to hold (i.e. to have non-zero bipartite entanglement
present) we need that $1\le k,l \le n-1$. This is the same condition as before when
the total state was pure. Thus, surprisingly, the condition for inseparability is
completely independent of temperature (although, two-site states do become separable
in the macroscopic limit even at zero temperature, as noted before).

We now look at the entanglement of the symmetric mixed state as a whole. Can we still
calculate the relative entropy of entanglement? This is in general very difficult to
do for multiparty mixed states, and some partial methods for upper bounds have only
been presented recently \cite{NJPhys}. We conjecture that the closest disentangled
state is now presumably the thermal average of the closest disentangled states for
individual $k$'s (this, I believe, is the same as the conjecture in \cite{Wei}, for
which Wei et al have offered a great deal of ``circumstantial evidence"; for example,
closest separable states have to possess the same symmetry as the entangled states for
which they minimise the relative entropy \cite{Werner}). I believe that this bound is
exact and that this can be proven using methods for calculating two site entanglement,
but I have not been able to show this yet. Even if this is not true, my method at
least gives us a very good upper bound which is sufficient to show how total
entanglement vanishes as $T$ becomes high. The relative entropy of entanglement
between these two states is given by (the right hand side of the inequality)
\begin{equation}
E (\sigma_T) \le \sum_k p_k \log p_k - \sum_k p_k \langle \Psi (k,n)| \log \left(
\sum_l p_l \rho_l \right) |\Psi (k,n)\rangle
\end{equation}
where $\rho_l$ is the closest disentangled state to the pure symmetric state
containing $l$ ones and $n-l$ zeroes. We have already seen that
\begin{equation}
\rho_l = \sum_{i=1}^l {l\choose
i}\left(\frac{k}{n}\right)^{l-i}\left(\frac{n-k}{n}\right)^{i} |\Psi (l,n)\rangle
\langle \Psi (l,n)| \; ,
\end{equation}
so that
\begin{eqnarray}
& E (\sigma_T) & \le  \sum_k p_k \log p_k - \nonumber \\
& - & \sum_k p_k \langle \Psi (k,n)| \log \left\{\sum_l p_l \sum_{i=1}^l {l \choose i}
\left(\frac{k}{n}\right)^{l-i} \left(\frac{n-k}{n}\right)^{i}\right\} |\Psi
(l,n)\rangle \langle \Psi (l,n)|
|\Psi (k,n)\rangle \nonumber \\
& = & - \sum_k p_k \log \sum_{i=1}^k {k \choose i} \left(\frac{k}{n}\right)^{k-i}
\left(\frac{n-k}{n}\right)^{i}\; .
\end{eqnarray}
The interesting conclusion here is the following. Suppose that we are at a high
temperature and that all symmetric states are equally likely, meaning that $p_k =
1/(n+1)$ for all values $k$ (basically, our state is an equal mixture of all symmetric
states). This, of course, in an approximation to the true density matrix, but it
becomes more and more accurate with the increase of temperature and it ceases to be
so when states other than symmetric become mixed in. The (upper bound to)
entanglement is then given by
\begin{equation}
E (\sigma_{T\rightarrow \infty}) \le \frac{1}{n+1} \log \sum_{i=1}^k {k \choose i}
\left(\frac{k}{n}\right)^{k-i} \left(\frac{n-k}{n}\right)^{i}
\end{equation}
The fraction inside the log tends to $n^2$ as $n$ becomes large, so that entanglement
scales as $\log n/n$. This is to be expected as entanglement grows as $\log n$ with
$n$, but the mixedness grows linearly with the number of state involved, $n+1$.
Therefore, in the thermodynamical limit, the overall mixed state entanglement also
disappears. This has to eventually happen, of course, if we believe that entanglement
is intimately linked with superconductivity and superconductivity also vanishes at
sufficiently high temperatures.

One kind of entanglement that we can say survives the thermodynamical high temperature
limit is the average of entanglements of individual symmetric states. This average
entanglement is given by
\begin{equation}
E_{avr} = \sum_k p_k E(|k,n-k\rangle) = \frac{1}{2}\sum_k p_k \log \frac{k(n-k)}{n} \;
.
\end{equation}
Note that if all probabilities go as $1/n$ - i.e. the symmetric state is maximally
mixed, then the entanglement scales as $\log n$ (the same as pure state at half
filling). This is expected, as there are $n+1$ states and each one has entanglement
proportional to $\log n$ and, so, on average, entanglement also goes as $\log n$.
However, this average entanglement, as we argued before is not a good measure as it requires
us to be able to address the symmetric states individually and
discriminate them from each other. This is not just difficult in practice, but is in
fact frequently even impossible in principle.

It is interesting to note that the ODLRO does survive the mixing of symmetric states.
Even when we have an equal mixture of all symmetric states the average ODLRO is given
by
\begin{equation}
\frac{1}{n+1} \sum_{k=0}^n \frac{k}{n}\frac{(n-k)}{n} = \frac{1}{2} -
\frac{1}{6}\frac{2n+1}{n} \rightarrow \frac{1}{6}
\end{equation}
where the arrow indicates the convergence when $n$ is large. Of course, at
sufficiently high temperatures the system will leave the subspace of symmetric states
and other states will also start to contribute. This eventually does lead to vanishing
of ODLRO, but the total entanglement and ODLRO may still disappear at different
temperatures. To calculate this exactly, we would need a much more detailed model and
a more extensive and careful calculation which lie outside of the scope of the present
paper. (Note: the same conclusions hold for the maximum singlet fraction in the
two-site density matrix which also survives the mixing in the thermodynamical limit;
this is, unfortunately and as pointed out before, not a suitable measure of
entanglement in our setting).

I conclude by showing that total correlations - quantum and classical - as quantified
by the mutual information \cite{Vedral1} can also easily be calculated for thermal
mixtures of symmetric states. Let us assume again that the symmetric states are maximally
mixed and each appears with the probability $1/(n+1)$. Then the mutual information is
given by
\begin{equation}
I = \frac{n}{n+1} \sum_k  \left(-\frac{k}{n}\log \frac{k}{n} - \frac{n-k}{n} \log
\frac{n-k}{n}\right) - \log(n+1) \; .
\end{equation}
For large $n$ this expression reduces to
\begin{equation}
I \rightarrow n - \log n \; .
\end{equation}
Since we know that thermal entanglement disappears in this limit, it is natural that
the mutual information is equal to the classical correlations and this then coincides with the
conclusion following eq. (\ref{mutual}).

\section{D-dimensional Symmetric States}

Extensions of all the considerations in this paper to $D$-dimensions are seen to be
very straightforward (similar generalizations to higher dimension symmetric states
were also considered in \cite{Wei}). We should actually be able to reproduce all the
above results in the generalized form, such that instead of qubits we have qutrits,
and so on. The generic symmetric state would now be written as:
\begin{equation}
|n_1,n_2,..n_d\rangle \label{higherspin}
\end{equation}
and it would be a totally symmetrized state of $n_1$ states $|1\rangle$, $n_2$ states
$|2\rangle$ and so on (it is also realistic to assume that the total number of
particles is conserved). This could, for example, represent higher spin fermions which
can occupy different lattice sites as in the rest of the paper. The closest state to
the one in eq. (\ref{higherspin}) in terms of the relative entropy is a mixture of the
states of the type
\begin{equation}
(\sqrt{n_1/N}|1\rangle + e^{i\phi} \sqrt{n_2/N}|2\rangle + ... e^{i (d-1)\phi}
\sqrt{n_d/N}|d\rangle)^{\otimes n} \; ,
\end{equation}
with the phase $\phi$ completely randomized as before. Knowledge of this closest state
allows us to compute the relative entropy of entanglement of any number of subsystems
of this system. All other results follow in exactly the same way. Nothing fundamental
is changed in higher dimensions which is why I will not say anything more on this
topic.

\section{Discussion and Conclusions}

In this paper I have analyzed the $\eta$ pairing mechanism which leads to eigenstates
of the Hubbard and similar models used in explaining high temperature
superconductivity. I have shown that they correspond to multi qubit symmetric states,
where the qubit is made up of an empty and a full site (two-electron spin singlet
state). I have also shown how to calculate entanglement and classical correlations for
such states. For pure states, entanglement of the total state increases at the rate
$\log n$ with the number of qubits $n$, while two-site entanglement vanishes at the
rate $1/n$. The two-site classical correlations, on the other hand, persist in the
thermodynamical limit. So, the ODLRO can be associated for pure states with total
entanglement or two-site classical correlations, but not with the two-site
entanglement. I have also demonstrated that the total entanglement for maximally mixed
symmetric states disappears at the rate $(\log n)/n$. Various mutual information
measures, which quantify the total amount of correlations in a given state, are also
computed and shown to be consistent with the calculations of classical and quantum
correlations.

There are many interesting issues raised by this work. Even if a consensus is reached
on the correct model for high $T_c$ superconductivity, and this is shown to contain
multipartite electron entanglement - which we have argued for in this paper - we are
still left with the question of being able to extract and use this entanglement. At
present there are no methods of extraction. Perhaps we can somehow extract electrons
from the superconductor and then use them for quantum teleportation or other forms of
quantum information processing.

It is presently believed that in order to perform a reliable and scalable quantum
computation we may need to be at very low temperatures, but the existence of high
temperature macroscopic entanglement may just challenge this dogma. Be that as it may,
I believe that the argument in favor of the existence of high temperature entanglement
does show that entanglement may be much more ubiquitous than is presently thought.
This may force us to push the boundary between the classical and the quantum world
towards taking seriously the concept that quantum mechanics is indeed universal and
should be applied at all levels of complexity, independently of the number or, indeed,
nature of particles involved.

\vspace{0.5cm}

\noindent {\bf Acknowledgements}. I would like to thank S. Bose, {\v C}. Brukner, H.
Fan, A. J. Fisher, J. Hartley C. Lunkes, C. Rogers, P. Scudo and T.-C. Wei for useful
discussion concerning this and related subjects. I am grateful to T.-C. Wei and H. Fan
for communicating their results to me prior to publication.

\end{document}